\documentclass[fleqn,usenatbib]{mnras}

\usepackage{newtxtext,newtxmath}
\usepackage[T1]{fontenc}

\DeclareRobustCommand{\VAN}[3]{#2}
\let\VANthebibliography\thebibliography
\def\thebibliography{\DeclareRobustCommand{\VAN}[3]{##3}\VANthebibliography}

\usepackage{graphicx}	% Including figure files
\usepackage{amsmath}	% Advanced maths commands
\newcommand{\msun}{\ensuremath{\mathrm{M}_{\odot}}}

\title[Binary AGN RHD]{Binary AGN simulations with radiation pressure reveal a new duty cycle, and a reduction of gravitational torque, through `minitori' structures}

\author[Williamson et al.]{
David J. Williamson,$^{1}$\thanks{E-mail: d.j.williamson@soton.ac.uk (DJW)}
Lars H. B\"{o}sch,$^{2,1}$
and Sebastian F. H\"{o}nig$^{1}$\thanks{E-mail: s.hoenig@soton.ac.uk (SFH)}
\\
$^{1}$School of Physics \& Astronomy, University of Southampton, Southampton, SO17 1BJ, UK\\
$^{2}$Institut f\"{u}r Theoretische Physik und Astrophysik, Christian-Albrechts-Universit\"{a}t zu Kiel, Leibnizstr. 15, 24118 Kiel, Germany\\
}

\date{Accepted 2021 December 21. Received 2021 November 22; in original form 2021 June 21}

\pubyear{2022}

\begin{document}
\label{firstpage}
\pagerange{\pageref{firstpage}--\pageref{lastpage}}
\maketitle

\begin{abstract}
We produce the first set of radiation hydrodynamics simulations of binary AGNs at parsec-scale separation in scale-model simulations. We use SPH for hydrodynamics, and raytracing to calculate optical depths and radiation pressure from the two AGNs. We confirm that, without radiation pressure, the sign of gravitational torque is sensitive to the binary parameters, although in one of our two orbital configurations the binary should coalesce in a time-scale of $<10^9$ yr. However, radiation pressure quickly destroys the `minitori' around each SMBH, drastically reducing gravitational torques and accretion, and greatly increasing the coalescence time-scale. Our simulations suggest a new `minitorus' duty cycle with a time-scale of $\sim10$ binary periods ($\sim10^6$ yr when scaling our models to a total binary mass of $2\times10^7$ M$_\odot$). The growth and blow-out phases of the `minitori' are of similar time-scales, and thus we expect about half of observed binary SMBHs to be active, in at least one component. The `minitorus' structure provides asymmetries that could be observed by infrared interferometry.
\end{abstract}

\begin{keywords}
quasars: supermassive black holes -- radiation: dynamics -- black hole mergers
\end{keywords}

%%%%%%%%%%%%%%%%%%%%%%%%%%%%%%%%%%%%%%%%%%%%%%%%%%

%%%%%%%%%%%%%%%%% BODY OF PAPER %%%%%%%%%%%%%%%%%%

\section{Introduction}\label{sec:intro}

Supermassive black holes (SMBHs) reside in the centre of almost every massive galaxy \citep{Kormendy1995, Richstone1998}. Through galaxy mergers, multiple SMBHs should exist in a single galaxy, at least for some time. The SMBHs of the parent galaxies should sink to the centre of the new galaxy through dynamical friction, form a binary, and perhaps merge, emitting gravitational waves \citep{Begelman1980,2014SSRv..183..189C}. While many merging galaxies can be observed, locating SMBHs during this phase is more complicated. In a morphologically complex merger remnant, the position of the SMBH is difficult to identify, and one SMBH or both SMBHs can be heavily obscured. Nevertheless, candidates for binary and dual SMBHs at different separations have been observed from kpc scales down to sub-parsec scales, e.g. via signatures of optical emission lines \citep[e.g.][]{Bogdanovic2009, Boroson2009, Dotti2009, Decarli2010, Eracleous2012, Comerford2013} or imaging in the radio \citep[e.g.][]{Rodriguez2006} or X-ray regime \citep[e.g.][]{Fabbiano2011}.\\

The frequency of observation is also related to the timescale for the whole merging process: The conservation of angular momentum and energy dictate that in each step in which the SMBHs approach each other the SMBH pair needs a steady transfer of angular momentum and energy from the SMBHs to the surrounding material, in order to decrease their separation. On large scales, this occurs on the dynamical time-scale of the galaxy ($\sim10^8$ yr), through stellar scattering and dynamical friction \citep{1943ApJ....97..255C,Begelman1980,2002MNRAS.331..935Y,2014SSRv..183..189C}. While there is some debate as to whether dynamical friction may be efficient in every galaxy \citep{1994MNRAS.271..317G,2017ApJ...840...31D,2018MNRAS.475.4967T}, clear coalescence mechanisms have been identified on these scales for a variety of parameters. Similarly, at very small separations ($\ll1$ pc), gravitational waves become an efficient source of torque, and the binary can quickly merge \citep{1964PhRv..136.1224P}.  However, at parsec-scale separations the amount of available stars for scattering is reduced (i.e. the `loss cone' is depopulated) such that neither scattering nor gravitational waves can merge the binary within the age of the universe. This is often referred to as the ``final parsec problem'' \citep{Milosavljevic2003}.\\

If binary black holes do merge, some mechanism must be found to supply significant negative torque to a binary with a separation of $\sim1$ pc. Propositions include (a) a third SMBH driving a merger through a complex 3-body orbit, (b) the `loss cone' being repopulated under certain circumstances, (c) a circumbinary gas structure driving coalescence through gravitational torques.

Models have found that introducing a third SMBH or other massive perturber can cause the binary to coalesce, potentially merging the third SMBH as well \citep{2006ApJ...651.1059I,2007MNRAS.377..957H,2018MNRAS.477.3910B}. However, these models have also found that this process depends strongly on SMBH masses, and often ejection of an SMBH is more likely than coalescence. If this is the dominant process, we should expect many galaxies to still host binary AGNs, depending on the rate at which additional SMBHs are supplied through mergers. If binary SMBHs are rare, then other processes must dominate.

Many simulations of dynamical friction and scattering assume idealised spherical galaxies, but relaxing this assumption may allow additional orbits that can repopulate the loss cone, providing more stars to be scattered, and allowing the SMBHs to coalesce \citep{2013ApJ...773..100K,2015ApJ...810...49V,2015ApJ...810..139H}. However, this mechanism depends strongly on the galaxy parameters and on numerical resolution, and it is still unclear if it can be a dominant process in the majority of mergers.

A number of simulations of binaries within a circumbinary gas disc have been performed, with a variety of smoothed-particle hydrodynamic codes \citep[e.g][]{1996ApJ...467L..77A,Escala2004,2005ApJ...630..152E,2006MNRAS.367..103D,2009MNRAS.393.1423C}, grid/mesh-based hydrodynamic codes \citep[e.g.][]{2002ApJ...567L...9A,2008ApJ...672...83M,2017MNRAS.466.1170M,2017A&A...604A.102T,2017MNRAS.469.4258T,2019ApJ...871...84M}, and semianalytic techniques \citep[e.g.][]{2008ApJ...679L..33K,2009PASJ...61...65H,2009MNRAS.398.1392L}. In these simulations, the binaries evacuate a cavity inside the circumbinary disc, and small accretion discs (`minidiscs') develop around each SMBH, with spiral accretion flows from the circumbinary disc to the minidiscs. It is generally found that the binary can drive gravitational instabilities in the disc, producing an asymmetric structure that can apply torque to the binary. Additionally, scattering of clouds in a clumpy disc can produce a dynamical friction effect in a similar way as with stars \citep{Roskar2015, Fiacconi2013}. However, depending on the disc viscosity and profile, binary properties, and hydrodynamic scheme,  the torque from either effect can be strong or weak, negative (decreasing binary separation) or positive (increasing binary separation). The gravitational torque is often dominated by gas in the minidisc close to the SMBHs, where numerical artefacts are most significant. Accretion from the disc onto the binary can supply further positive torque, potentially countering any negative torque from an asymmetric disc. The relative importance (and sign!) of these effects is not yet constrained, and further simulations are still required.

Additionally, few of these models are specialised SMBH models. Instead, many are generic binary models and are applied to the full range of scales, from binary stars to binary SMBHs. In particular, binary SMBHs are likely to be active galactic nuclei (AGNs), due to gas inflow triggered by the galaxy merger, and should be strongly affected by radiation pressure from the luminous AGNs. These simulations also tend to focus on binary SMBHs at small separations (e.g. $\sim100$s of $R_g$), where observational signatures can have time-scales of $t<10^2$ yr, and are easier to detect.

In the standard model of unification of active galactic nuclei (AGN) \citep[e.g.][]{Antonucci1985, Antonucci1993, Urry1995, RamosAlmeida2017}, an AGN consists of an SMBH, surrounded by a hot accretion disc, in turn surrounded by a geometrically thick ``torus'' of obscuring dusty gas. Recent high-angular resolution observations in the infrared (IR) \citep{Hoenig2012, Hoenig2013, Tristram2014, LopezGonzaga2014, LopezGonzaga2016, Leftley2018} reveal a somewhat more complex structure, consisting of an equatorial thin disc on sub-parsec scales and a polar-elongated feature which can even be observed on scales of 100s of $\mathrm{pc}$ \citep{Asmus2016}. This is interpreted as a dusty wind driven from the ``torus'' by radiation pressure from the central UV/optical continuum emission of the AGN acting on the dust, as modeled in several radiation-hydrodynamics (RHD) simulations \citep[e.g.][]{Dorodnitsyn2012, Wada2012, Dorodnitsyn2017, Chan2016, Chan2017, Williamson2019}. This is different from line-driven winds that primarily act on the dust-free gas in the broad-line region \citep[e.g.][]{2000ApJ...543..686P,2016MNRAS.458..293M}.

Recent ALMA observations have elaborated on this complex structure, revealing a complex structure of superimposed rotation and outflows, and even apparent counter-rotation \citep{2016ApJ...823L..12G,2018ApJ...859..144A,2018ApJ...867...48I,2019A&A...628A..65A,2019ApJ...884L..28I,2019A&A...632A..61G,2019A&A...629A...6T}. On this scale, galaxy physics such as supernova feedback also become important \citep{Wada2012}. This complex dynamical picture can not be modelled by purely gravitating SMBHs within a flat circumbinary disc.

A key factor is that radiative effects become increasingly important in binary SMBHs. Radiation pressure from an AGN accretion disc could reduce the accretion rate (reducing positive torque), stabilise against instabilities or induce different instabilities in the disc (or a torus), and drive an outflow that can carry angular momentum. However, no binary SMBH model has included radiation pressure from AGN accretion discs. Hence, while radiation pressure can have a significant effect on black hole coalescence, whether the net effect is positive or negative is yet unclear.

Binary AGNs are presently of particular interest in the era of multi-messenger physics \citep{2019NewAR..8601525D}. Gravitational wave detectors such as the upcoming LISA \citep{2005MNRAS.361.1145R} will provide constraints on the merger rates of SMBHs. Upcoming high-resolution infrared imagers such as GRAVITY+ \citep{gravplus} will resolve the inner regions of AGNs, potentially revealing signatures of binarity. Realistic numerical models of binary AGNs can identify characteristic features of binaries, to interpret and guide these future observations.

In this paper, we perform RHD simulations of binary SMBHs of parsec-scale separation within a torus-scale circumbinary disc, taking into account the effects of radiation from SMBH accretion discs, including photoionisation, heating, and radiation pressure onto the circumbinary material. In doing so, we will attempt to determine:

\begin{enumerate}
\item whether radiative effects help or hinder binary SMBH coalescence at $\sim$parsec separations
\item if radiative effects develop or mask signatures that can be used to identify binary AGNs
\end{enumerate}

\section{Method}

Our simulations build upon the AGN RHD model of \citet{Williamson2019}, which uses the public version of the N-body+hydrodynamics code GIZMO \citep{Hopkins2015} in pressure smoothed-particle hydrodynamics (P-SPH) mode to model the toroidal structure around a single SMBH as a wind driven by a strong radiation pressure. Hence, we will only briefly describe the main characteristics of the code, and the new modifications, and refer to the underlying publications \citep{Williamson2019,Hopkins2015} for more detailed information.

\subsection{Gravity, orbits, and accretion}

Self-gravity is included for the gas particles, using a Barnes-Hut octtree \citep{Barnes1986}. Gravity from the SMBHs is modelled as a softened Keplerian,
\begin{equation}
    \vec{g}_\mathrm{SMBH} = -G\left(\frac{M_1}{r_1^2+c_\mathrm{BH}^2}\,\vec{e}_{r1} + \frac{M_2}{r_2^2+c_\mathrm{BH}^2}\,\vec{e}_{r2}\right)
\end{equation}
where $r_1$,$r_2$ are the distances between the SMBHs of mass $M_1$ and $M_2$ and the particle, $\vec{e}_{r1}$ and $\vec{e}_{r2}$ are the corresponding unit vectors and $c_\mathrm{BH}=10^{-5}$~pc is the smoothing length. When the distance from a particle centre to the SMBH is within $10^{-3}$ pc plus the particle's smoothing length, it is instantly accreted onto that black hole\\

As the gas is self-gravitating and allowed to cool, it can collapse into dense cores, which should collapse to form stars. These dense structures also significantly slow down the simulation. Hence we apply a star formation `sink', stochastically deleting particles with a density above a threshold density of $n_H=10^{10}$ at a rate inversely proportional to the particle's freefall timescale.

The orbits of the SMBHs are calculated analytically, as idealised two-body Keplerian elliptical orbits. We track the mass, momentum, and angular momentum accreted onto the SMBHs, and the gravitational forces and torques applied from the gas to the SMBHs. However to prevent numerical effects from destabilising the system, we do not change the SMBH orbits or their gravitational masses through the simulation. As the cumulative torque and accretion over our simulation time is low, this approximation is reasonable.

\subsection{Radiative transfer and thermodynamics}

\subsubsection{Application of radiation}

As in previous work \citep{Williamson2019}. heating and cooling rates, radiation pressure strength, opacity (which is typically dust dominated, as temperatures are typically $T_\mathrm{gas}<1000$ K), and other terms used in post-processing such as dust temperature ($T_\mathrm{dust}$), are pretabulated using the photoionization code Cloudy \citep{Ferland2013, Ferland2017}. Here, the radiation-dependent properties of a particle $p$ are a function of four parameters - the temperature~$T$ and density~$\rho$ of the gas particle, the optical depth $\tau$ from the AGN to the particle, and unextinguished flux of the AGN $F$ at the particle's location. $\rho$ and $T$ are determined from the gas particle properties in the simulation, and the derivation of $F$ and $\tau$ in a binary AGN model are explained in Section~\ref{sec:rt}.

The Cloudy models are performed at a fixed gas density and temperature (which is not generally equal to the dust temperature)), and the heating and cooling rates extracted are assumed to be instantaneous rates. These are applied to the internal energy of the particle, with the timestep criterion that a particle can not reduce its internal energy by more than $10\%$ or increase its internal energy by $100\%$ in a single timestep.

\subsubsection{Sources of radiation}

In our binary system both SMBHs are sources of radiation. The corresponding luminosities~$L_{i}$ for each SMBH with the masses~$M_{i}$ are calculated through
\begin{equation}
    L_{i}=\gamma_\mathrm{Edd} L_{\mathrm{Edd},i}=1.26\times 10^{44}\,\gamma \frac{ M_{i}}{10^6 \msun}\,\mathrm{erg}\,\mathrm{s}^{-1}
\end{equation} 
with a constant Eddington factor $\gamma_\mathrm{Edd}=0.05$ for both SMBHs. While a luminosity depending on the actual accretion rate is beyond the scope of this model, the chosen value for the Eddington factor is within the range expected for the $M\sim10^6\msun$ AGNs considered in this work \citep{2014ApJ...791..113D,2015ApJ...815..129S}. 

As in \citet{Williamson2019}, the unextinguished flux of an AGN is assumed to be azimuthally isotropic but anisotropic in polar angle because the source of the radiation is a geometrically thin disc. The flux of each SMBH is described as
\begin{equation}
    F_{i}(r_{i},\theta_{i}) = \frac{L_{i}}{4\pi r_{i}^2}\,f(\theta_{i}),
    \label{eq: Flux}
\end{equation}
where $r_{i}$ is the distance from each of the SMBH, $\theta_{i}$ is the angle between the polar axis and the line to the corresponding SMBH and $f(\theta_{i})$ is an anisotropy function. We have modified the anisotropy function of  \citet{Netzer1987} so that the equatorial flux is non-zero. This models the contribution of isotropic emission from an x-ray halo, and deviations from a perfect accretion disc such as warping or accretion disc winds. We define the anisotropy function as
\begin{equation}
    f(\theta)=\frac{1+a\cos\theta+2a\cos^2\theta}{1+2a/3},
\end{equation}
where $a=(\eta_a-1)/3$ depends on the anisotropy factor $\eta_a$, equal to the ratio between polar flux and the equatorial flux and fixed to $\eta_a=100$ in this work. 

\subsubsection{Radiation transfer from two AGNs}\label{sec:rt}

We calculate the optical depth towards each particle from each of the SMBHs independently. For runs that don't include radiation pressure, this optical depth is assumed to be large, and set to $\tau=7$. In runs that include radiation pressure, we use the raytracing algorithm described in \citet{Williamson2019}, using particle opacities calculated from Cloudy. In this previous work, only one AGN was used, and the radiation-dependent properties of a particle $p$ are a function of four parameters. However, with two AGNs, we get a six-dimensional parameter space as we calculate two AGN-to-particle optical depths $\tau_{1,2}$ and two unextinguished AGN fluxes $F_{1,2}$ for each particle.

A six-dimensional table would be prohibitively large, so we use an approximation to model the radiation field using the existing four-dimensional tables. We can expect that for nearby particles the radiation field is typically dominated by only one SMBH and the other one adds only minor perturbations, and for distant particles, the two SMBHs act similarly to a single large SMBH, again with minor perturbations. Therefore we calculate an effective flux and effective optical depth through weighted averages,
\begin{align}
    \frac{1}{\tau_{\mathrm{e}}}&=\frac{F_1/\tau_1 + F_2/\tau_2}{F_1+F_2}\\
       F_{\mathrm{e}}&=F_1 e^{\tau_{\mathrm{e}}-\tau_1} + F_2 e^{\tau_{\mathrm{e}}-\tau_2},
\end{align}
Based on this parameter space $(T,\rho,F_{\mathrm{e}},\tau_{\mathrm{e}})$ we use the same tables as \citet{Williamson2019} for the calculation of the heating~$\tau_{\mathrm{heat}}$ and cooling rates~$\tau_{\mathrm{cool}}$ and the magnitude of the radiation acceleration~$a_{\mathrm{rad}}$. The direction of radiative acceleration~$\vec{a}_{\mathrm{rad}}$ is taken as a weighted average of the unit vectors $\hat{r}_{i}$ pointing from each SMBH $i$ to the particle $p$, specifically
\begin{equation}
    \vec{a}_{\mathrm{rad}}= a_{\mathrm{rad}} \frac{\hat{r}_{1} F_1 e^{\tau_{\mathrm{e}}-\tau_1} + \hat{r}_{2} F_2 e^{\tau_{\mathrm{e}}-\tau_2}}{F_1 e^{\tau_{\mathrm{e}}-\tau_1} + F_2 e^{\tau_{\mathrm{e}}-\tau_2}}
\end{equation}
In this description it is even possible that the radiation pressure of both of the SMBHs cancel each other out. Nevertheless the heating and cooling still remains unaffected by this.

To calculate $\vec{a}_{\mathrm{rad}}$, we use $e^{\tau_\mathrm{e}-\tau_i}$ terms even when the $e^{\tau_\mathrm{e}}$ terms cancel out, to avoid rounding errors. We also include a $\tau$ cutoff at $\tau_c=7$, beyond which radiative acceleration becomes negligible. The above procedure can, in some circumstances, scale up both $\tau_\mathrm{e}$ and $F_\mathrm{e}$, especially if the closest AGN is heavily extinguished. To avoid extreme values and rounding errors when $\tau_\mathrm{e}>\tau_c$, in these cases we scale the unextinguished flux by a factor of $e^{\tau_c-\tau_\mathrm{e}}$ and set $\tau_\mathrm{e}=\tau_c$.

\subsection{Simulations \& Scaling}

\begin{table}
	\centering
	\caption{Units for scaled runs. From top to bottom the terms are: the name of the model; $a$, the length unit and maximum binary separation; $r_s$, the sublimation radius; $T$, the time unit and binary period; $v$, the velocity unit; $M_t$, the total binary mass. First column shows units used in the simulation, second column shows units scaled up to a total SMBH mass of $2\times10^6$ \msun}
	\label{tab:units}
	\begin{tabular}{rll}
	    Unit & Simulation & Scaled\\
		\hline
		$a$ & $0.035$ pc & $1.11$ pc\\
		$r_s$ & $0.0012$ pc & $0.037$ pc\\
		$T$ & $13.7$ kyr & $77$ kyr\\
%		$v=a/T$ & $2.50$ km/s & $$ km/s\\
		$M_t=M_1+M_2$ & $2\times10^3$ \msun & $2\times10^6$ \msun\\
		\hline
	\end{tabular}
\end{table}

\begin{table*}
	\centering
	\caption{Summary of simulations. From left to right the columns are: the name of the simulation; the initial conditions for the simulation, including the time of the snapshot used for initial conditions if the initial conditions have been extracted from a previous simulation; the eccentricity of the binary orbit (where applicable); the mass ratio of the two SMBHs (where applicable); whether radiation pressure is active; and the total run time of the simulation, in units of the binary orbital period.}
	\label{tab:runs}
	\begin{tabular}{rlllll}
	    Name 				& Initial Conditions 		& $e$	& $M_1/M_2$ 	& Radiation Pressure & Run time\\
		\hline
		norad\_ecc		& Binary ICs			& 0.5		& 3			& N				& 99.9T\\
		rad\_ecc\_early		& norad\_ecc, $t=14.7T$	& 0.5		& 3			& Y				& 9.3T\\
		rad\_ecc\_late		& norad\_ecc, $t=99.9$	& 0.5		& 3			& Y				& 5.0T\\
		\hline
		norad\_circ		& Binary ICs			& 0.0		& 1			& N				& 99.9T\\
		rad\_circ\_early		& norad\_circ, $t=15.2T$	& 0.0		& 1			& Y				& 8.1T\\
		rad\_circ\_late		& norad\_circ, $t=99.9$	& 0.0		& 1			& Y				& 5.0T\\
		\hline
		norad\_single		& Single ICs			& N/A	& N/A		& N				& 17.1T\\
		rad\_single	& norad\_single, $t=9.5T$	& N/A	& N/A		& Y				& 2.8T\\
		\hline
	\end{tabular}
\end{table*}

We perform $2$ sets of binary simulations and one set of simulations of a single SMBH. Each set consists of runs in two phases. The first phase consists of one run without radiation pressure, and with simplified radiation transfer. The second phase consist of one or more continuation runs that include our full radiative transfer model. In both cases, heating and cooling rates are calculated with CLOUDY, but in the first phase all particles are assumed to have an optical depth of $\tau=7$. We use this approach to equilibrate the system over many orbits prior to activating radiation pressure, and to provide a basis to show the effects of radiation pressure, while retaining a realistic gas equation of state even when radiation pressure is switched off. This can also be considered a simple model for an AGN `duty cycle', where the system passes through a stage of accretion, followed by an active `blowout' stage (see section~\ref{sec:duty}).

One set of binary simulations considers an equal mass SMBH binary in a circular orbit, while the other considers an SMBH binary with a 3:1 mass radio in an orbit with an eccentricity of $e=0.5$. Thus we investigate the circumbinary torus torque in the simplest possible binary orbit configuration, as well as investigate the effects of an asymmetric binary on disc instabilities and the resulting disc torque. This is an exploratory work, where we identify the general phenomena of each configuration, and we do not perform a full parameter search with a large number of simulations.

A very fine mass resolution is required to resolve wind generation from radiation pressure \citep{Williamson2019}. However, we also want the circumbinary torus mass to be comparable to that of the black holes, so that the gravitational torque is significant. A large mass and a fine mass resolution require a large number of particles. This can produce very computationally expensive simulations.

To reduce this expense, we perform scaled simulations. These can be calculated more quickly, as simulation timesteps are typically dominated by size-invariant thermodynamic timesteps such as the cooling timestep. Scaling down the system reduce the dynamical time, without reducing the cooling time, allowing the system to evolve further over the same number of cooling timesteps without the numerical instabilities that would be introduced by increasing the cooling timestep. This is analogous to the `reduced speed of light' approximation, which similarly reduces the dynamical time to be closer to the more dominant radiation timestep, and will similarly have some effect on the evolution of thermodynamic instabilities.

Hence we perform scaled simulations with a total binary SMBH mass of $2\times10^3$ \msun, and a gas disc of $10^7$ particles with a mass of $10^3$ \msun. Here the binary:disc mass ratio is 2:1. We consider these simulations to be scaled down in mass by a factor of $1000$, and to represent a total binary SMBH mass of $2\times10^6$ \msun.

To produce the scaled simulation, we set the ratios of distances (binary maximum separation, sublimation radius, circumbinary torus inner radius) and the Eddington factor to typical full-scale values. Although the actual distances where sublimation occurs depends on inclination and optical depth, we define `the' sublimation radius $r_s$ through $L/(4\pi r_s^2)=I_\mathrm{sub}$, where $I_\mathrm{sub}=10^{7.58}$~erg s$^{-1}$ cm$^{-2}$ is the intensity where half of the dust is destroyed by the unextinguished flux from a single AGN, according to our Cloudy tables. This gives $r_s=0.037 (\gamma_\mathrm{Edd}/0.05)^{1/2} (M/10^6)^{1/2}$ pc. As the sublimation radius scales as $r_s\propto (\gamma_\mathrm{Edd} L_\mathrm{Edd})^{1/2} \propto (\gamma_\mathrm{Edd} M)^{1/2}$, we scale down all distances by a factor of $1000^{1/2}\approx32$. The orbital period at the inner disc boundary follows Kepler's laws, as does the binary orbital period, both proportional to ${a^{3/2}M^{-1/2}}\propto M^{1/4}$. This means the ratio between the inner disc orbital period and the binary orbital period does not depend on scaling, as both are scaled down by a factor of $1000^{1/4}\approx5.6$.

To quantitatively compare our models, we present results using binary semi-major axis as a length unit $a$, and binary orbital period as a time unit $T$. Physical values for these units are given in Table~\ref{tab:units}. Angular momentum is given in units of $J_0$, the initial angular momentum of the Keplerian binary, which differs slightly between the two orbit models. We also quote the sublimation radius.

We set the surface density profile of the circumbinary torus by forcing the Toomre parameter $Q=1.2$ at all radii, so that the disc is borderline stable against gravitational instabilities. As a consequence instabilities may be driven by the binary, but the disc will not rapidly collapse and fragment under its own self-gravity. This gives an outer radius of $R_o\sim23a$. The inner radius of the disc in binary simulations is $1.43a$, which is a little smaller than the cavity size eventually produced by the binary. In the single SMBH model, the inner disc radius is $0.03$a.

In the first phase, we run the models without raytracing and radiation pressure until $t=100T$, though as stated above, some radiation transfer is still processed, as gas heating and cooling rates are calculated using the CLOUDY tables, under the assumption that the gas column to the AGN is very optically thick, with $\tau=7$. In this phase, snapshot dumps are produced every $\Delta t=0.1T$ until $t=2T$, to confirm the simulations are progressing and equilibrating correctly, and from then every $\Delta t\sim19T/20$, to investigate the long term evolution without using a prohibitively large amount of disc space. The time between dumps is slightly less than $T$, so that we can see snapshots of the system at different phases of its orbit. We label these runs norad\_ecc, and norad\_circ. For the single SMBH run, norad\_single, we terminate the run at $t\sim6T$, as the system equilibrates more quickly.

Radiative transfer is more computationally expensive, and the model can only be evolved for a smaller number of orbits. We start one pair of binary models using the final conditions of the runs without raytracing. We label these runs rad\_ecc\_late and rad\_circ\_late.  However, by $t=100T$, much of the gas has been lost from the system through accretion. We therefore also perform a pair of models using the conditions at $t\sim15T$. For these four continuation runs, we take snapshots every $0.1T$, and terminate the run after the central cavity is completely evacuated, within $\sim5-10T$. We label these runs rad\_ecc\_early and rad\_circ\_early. For the single SMBH, we produce a single run with radiation pressure that continues from the end of norad\_single, and we label this rad\_single.

The simulation names, initial conditions, and parameters are summarised in Table~\ref{tab:runs}.

\section{Results}

\begin{figure}
\begin{center}
\includegraphics[width=\columnwidth]{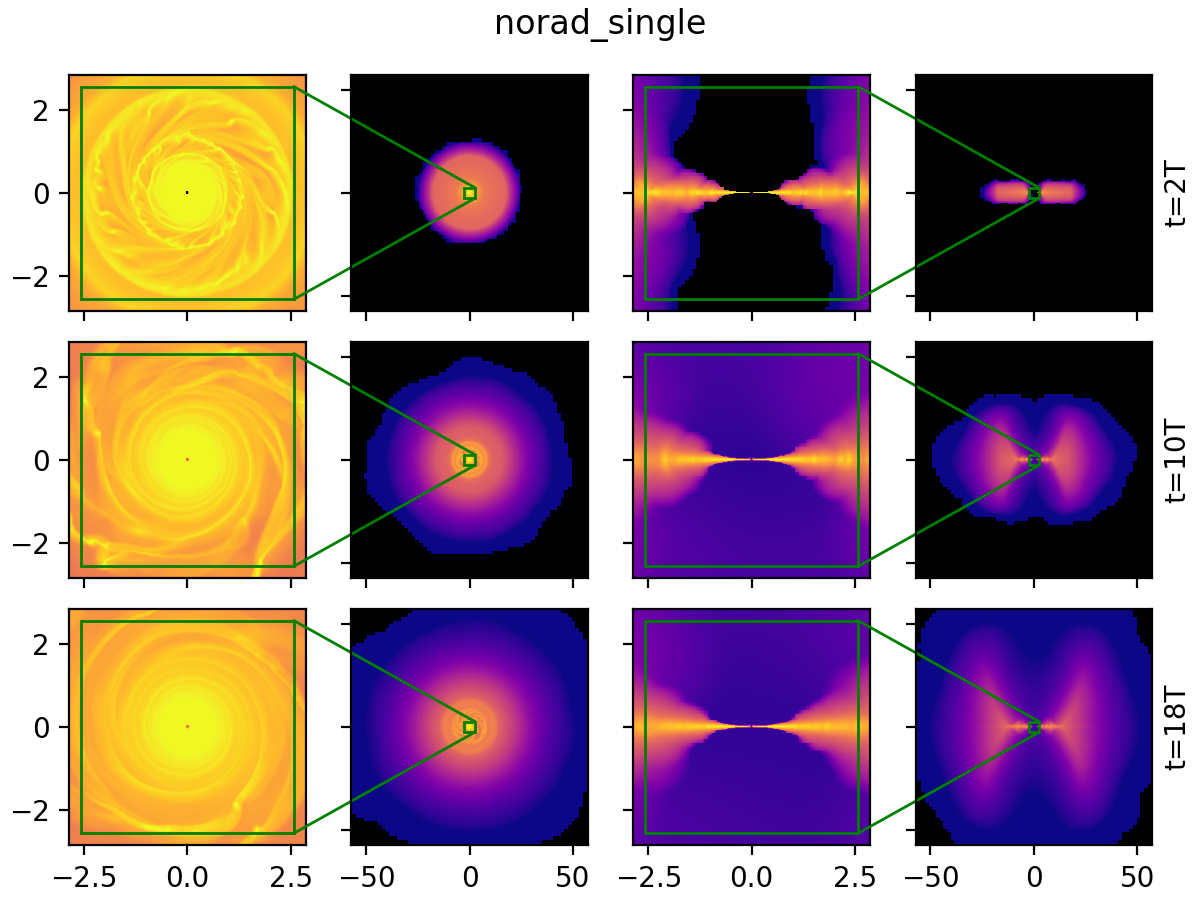}
\includegraphics[width=\columnwidth]{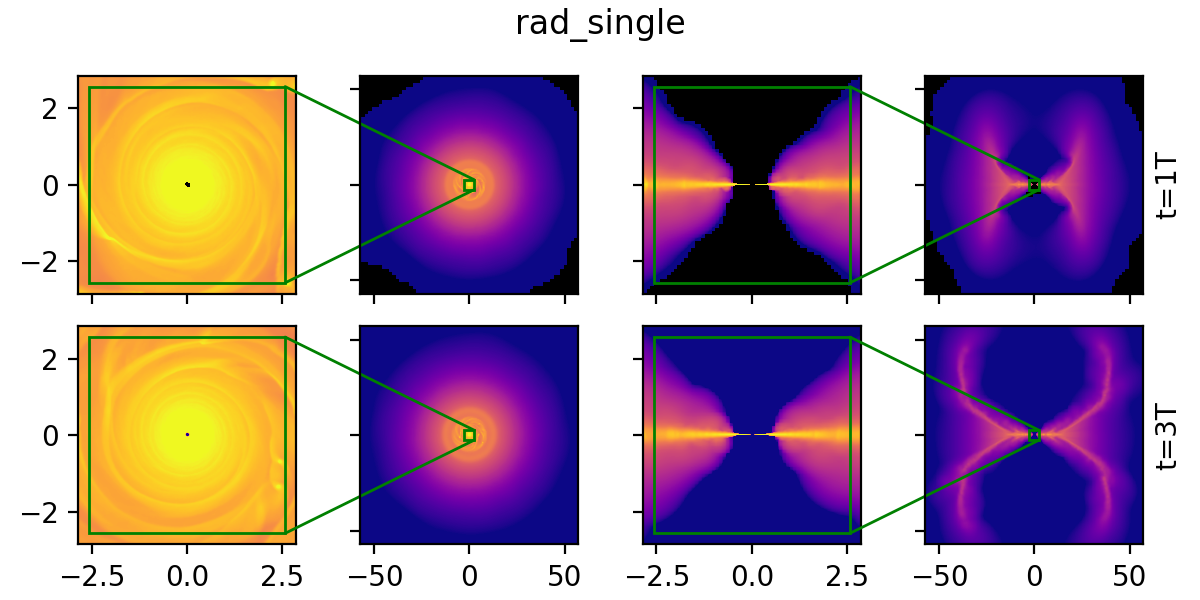}
\end{center}
\caption{\label{unary_evol}
Density slices of evolution of single SMBH model. Length units are the semi-major axis of the binary, $a$.}
\end{figure}

\begin{figure}
\begin{center}
\includegraphics[width=\columnwidth]{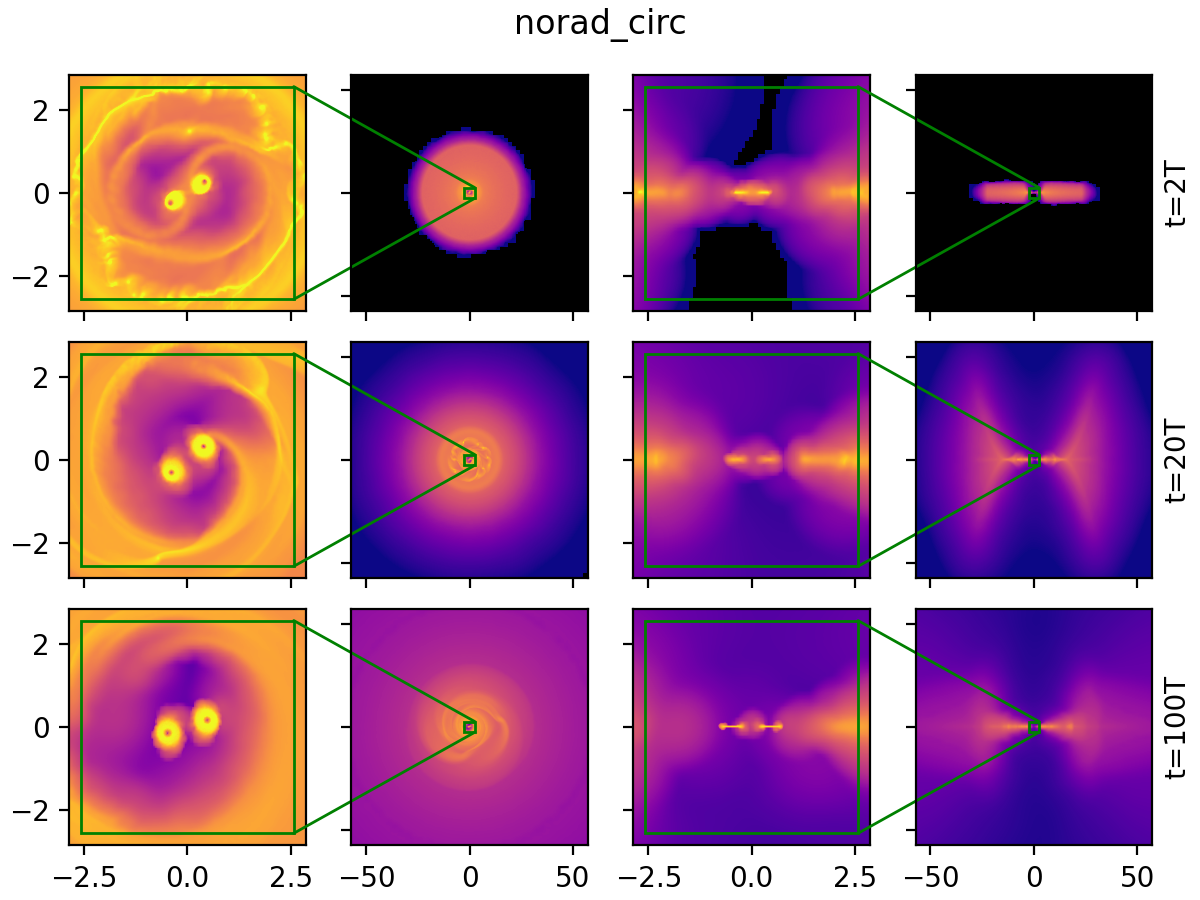}
\includegraphics[width=\columnwidth]{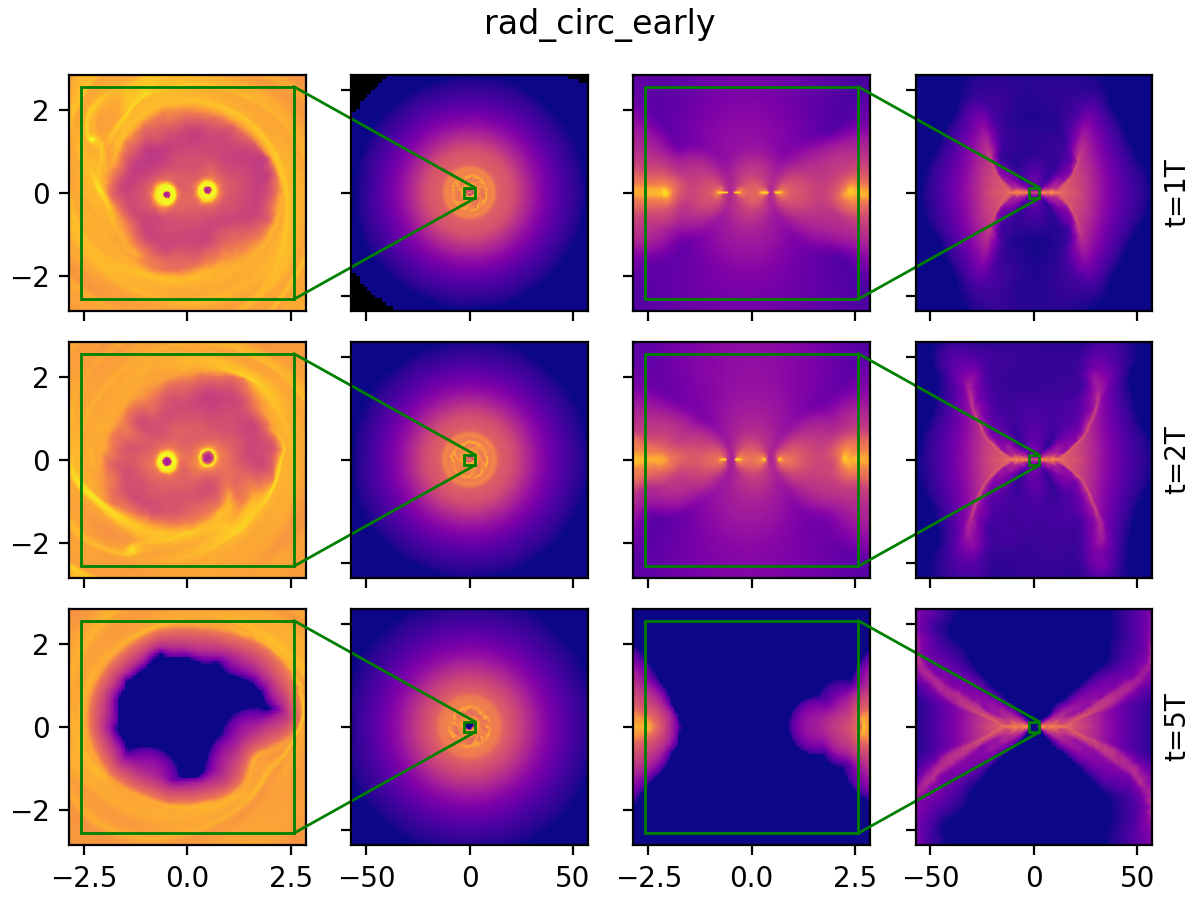}
\includegraphics[width=\columnwidth]{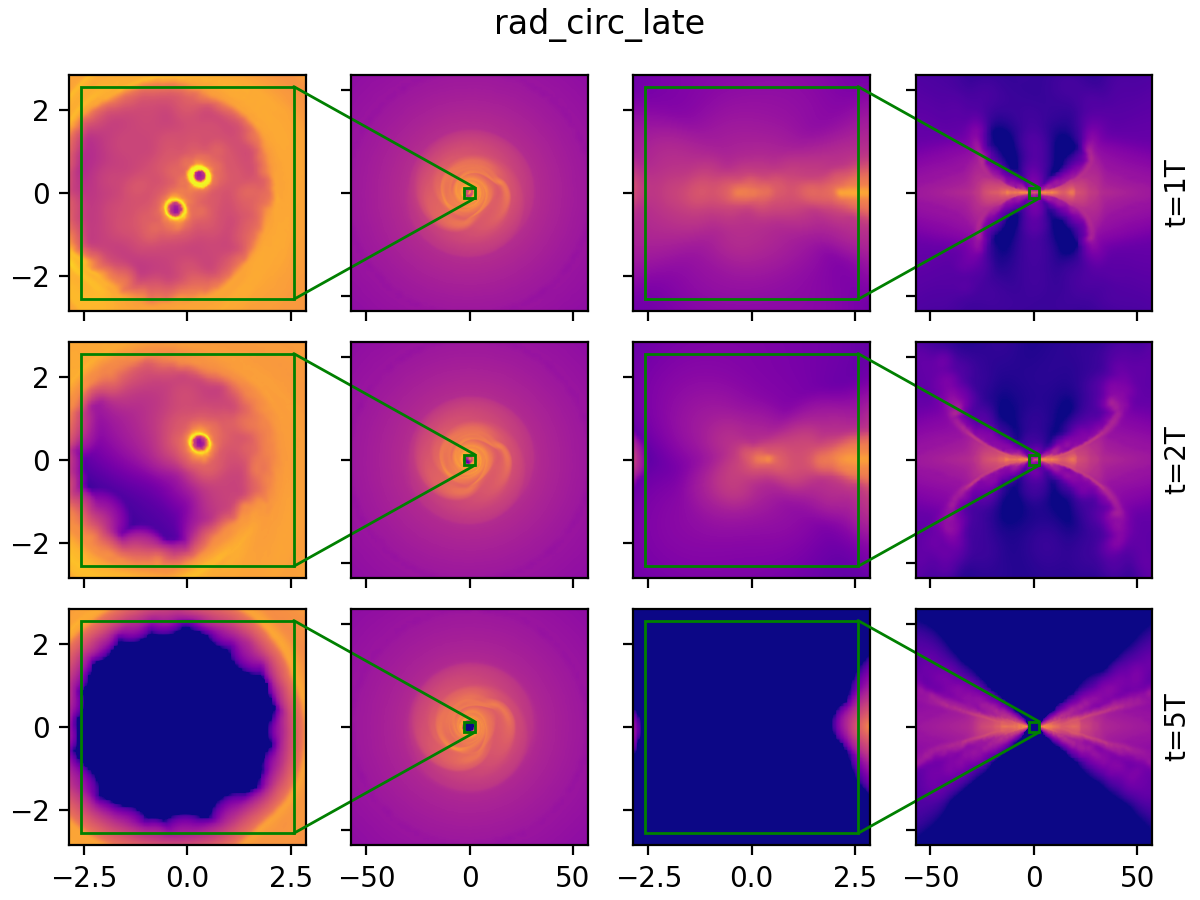}
\end{center}
\caption{\label{circ_evol}
Density slices of evolution of circular binary model. Length units are the semi-major axis of the binary, $a$.}
\end{figure}

\begin{figure}
\begin{center}
\includegraphics[width=\columnwidth]{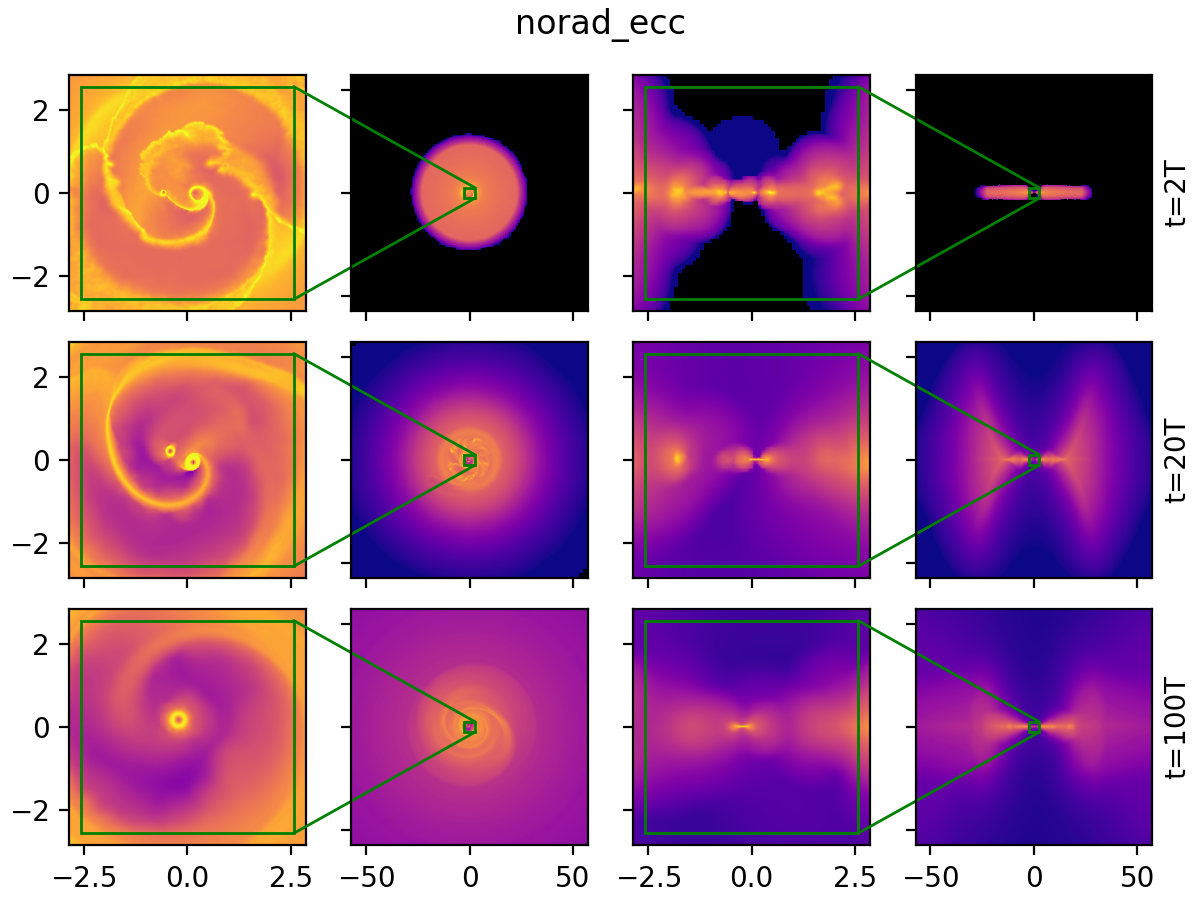}
\includegraphics[width=\columnwidth]{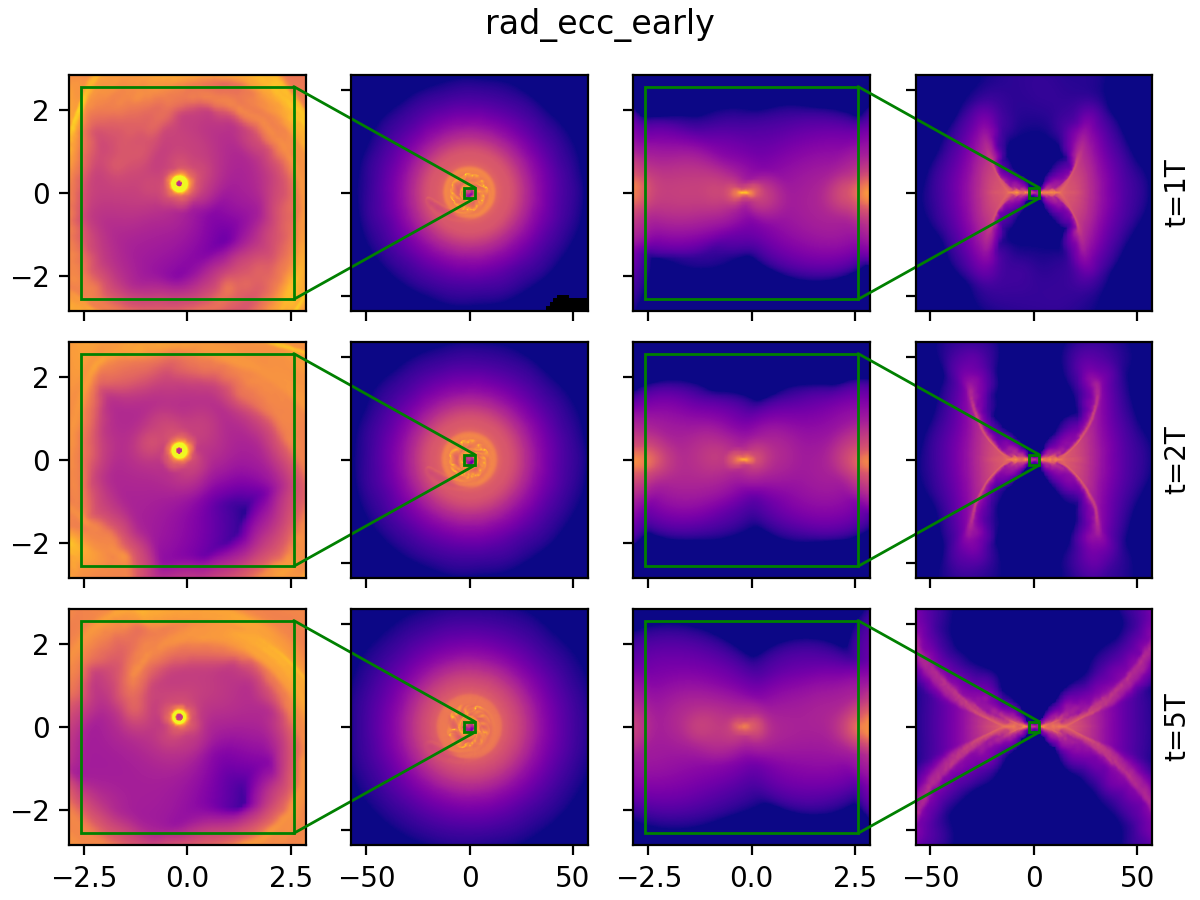}
\includegraphics[width=\columnwidth]{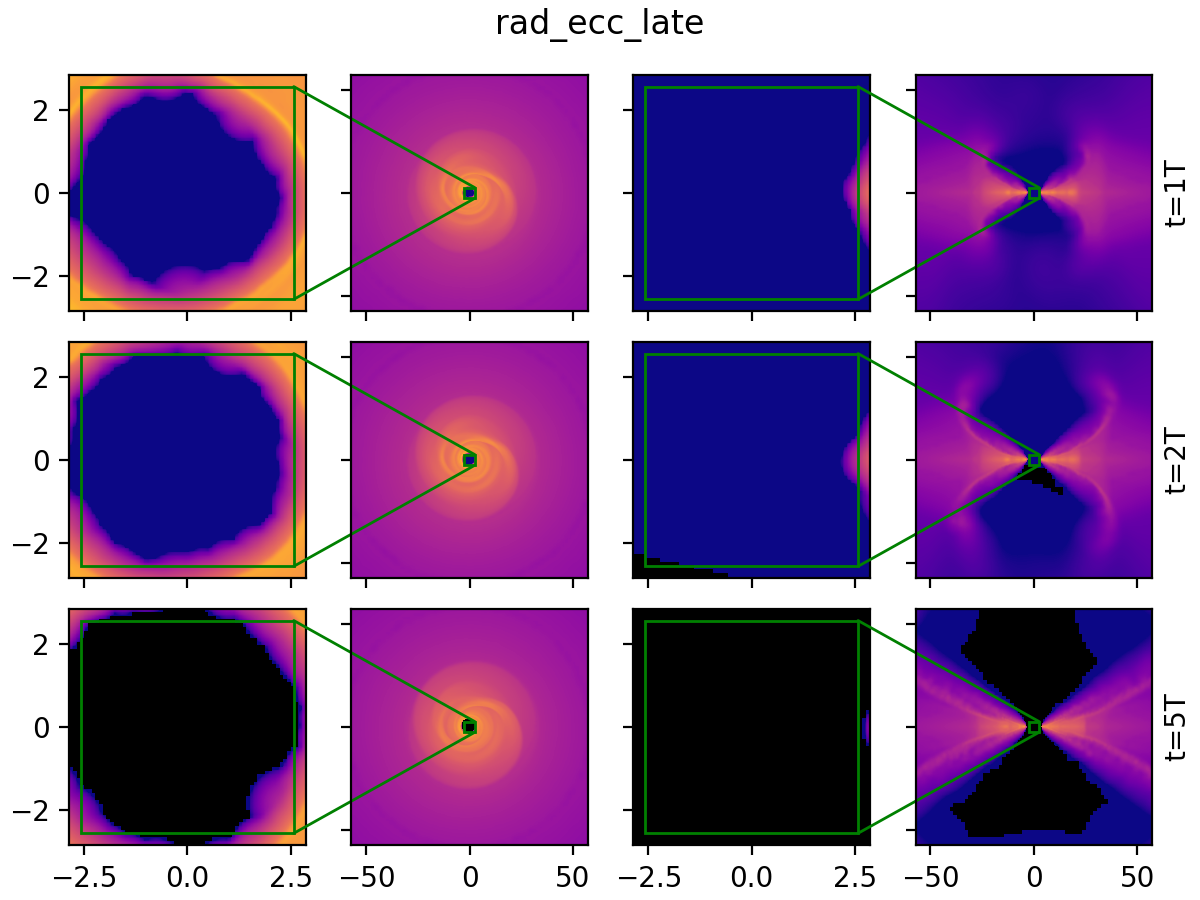}
\end{center}
\caption{\label{ecc_evol}
Density slices of evolution of elliptical binary model. Length units are the semi-major axis of the binary, $a$.}
\end{figure}

\begin{figure}
\begin{center}
\includegraphics[width=\columnwidth]{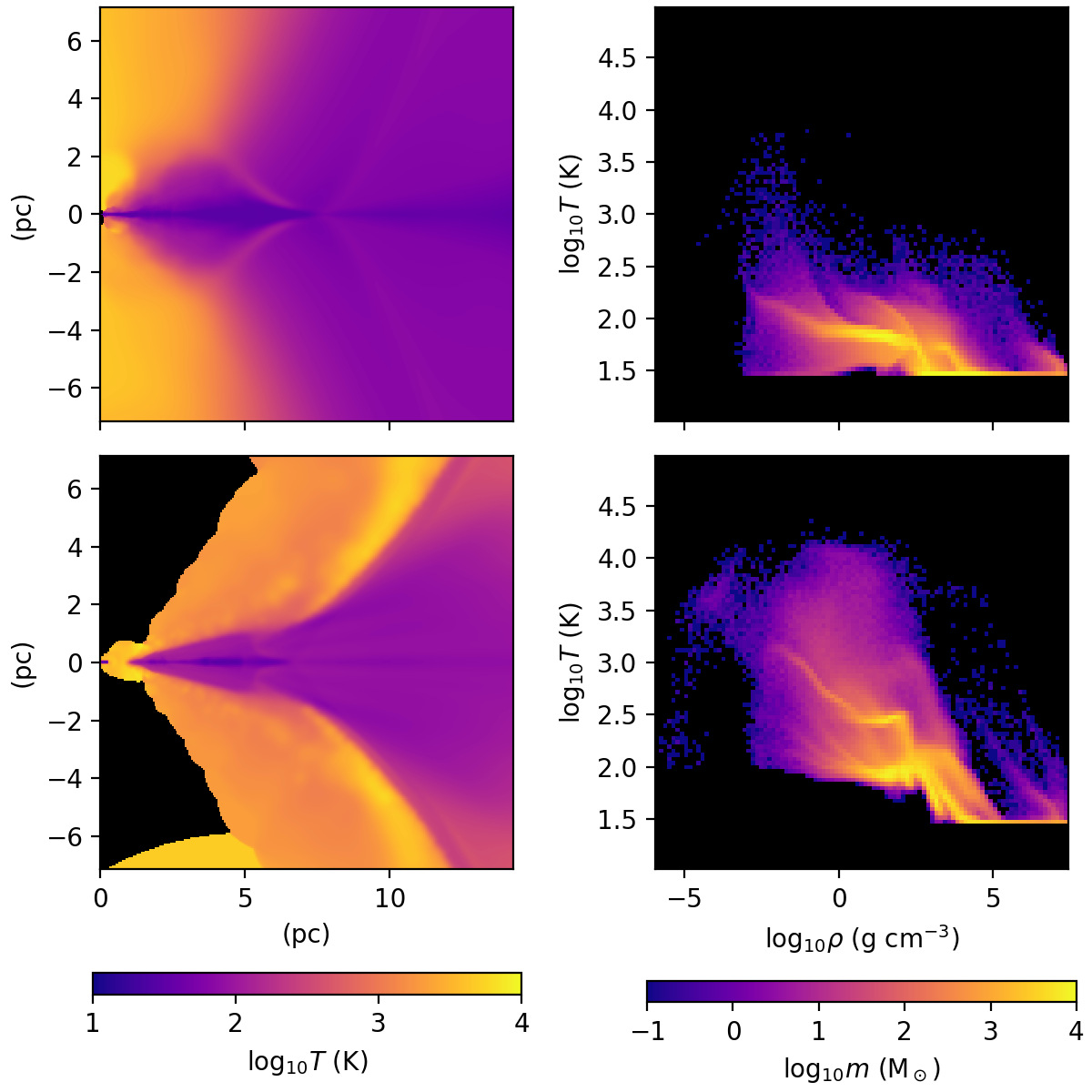}
\end{center}
\caption{\label{tempplot}
An illustration of typical temperature fields in the simulations. Left: azimuthal mean temperatures. Right: $\rho-T$ phase plot. Top: norad\_ecc at $T=20$. Bottom: rad\_ecc\_early at $T=2$. 
}
\end{figure}

\begin{figure}
\begin{center}
\includegraphics[width=\columnwidth]{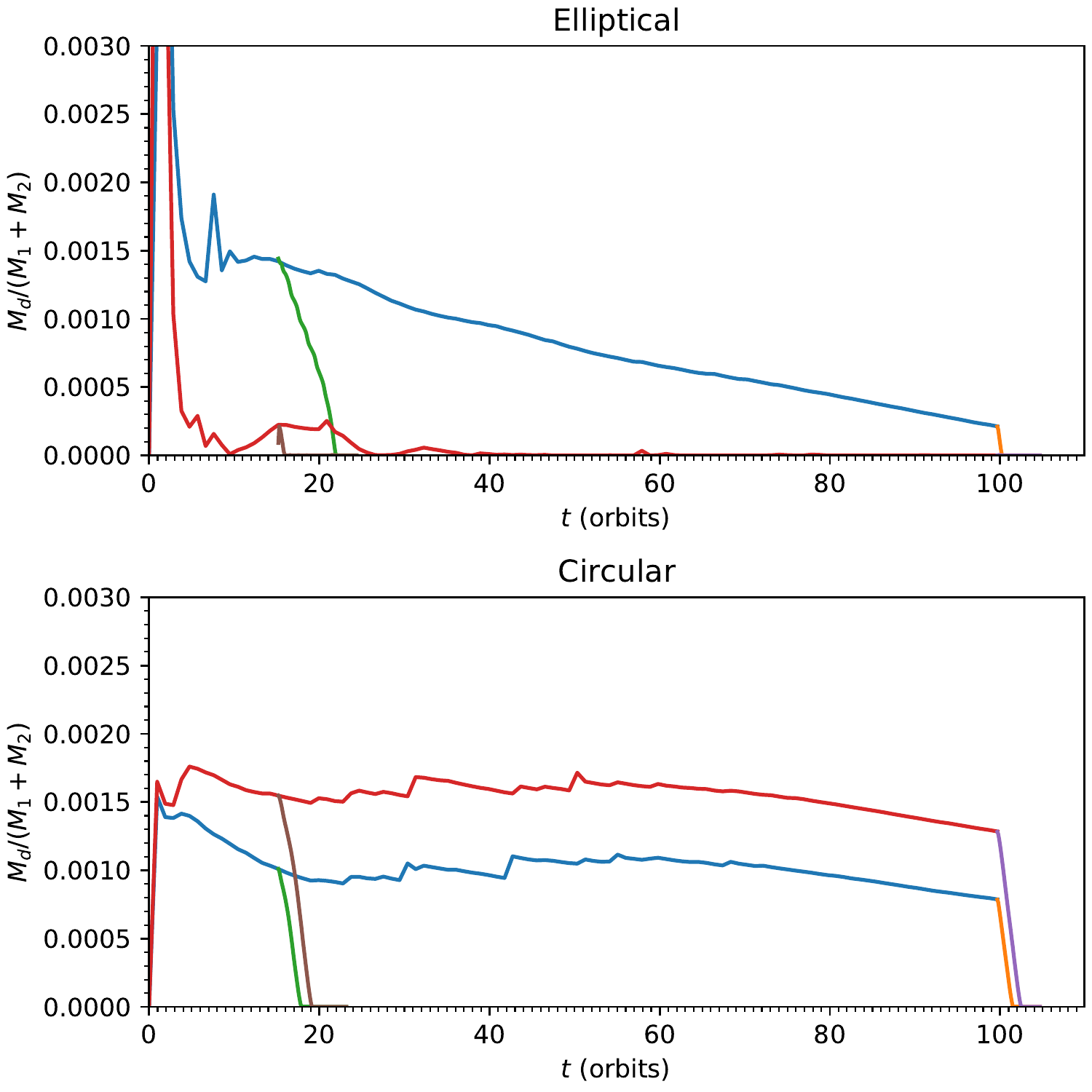}
\end{center}
\caption{\label{windgen}
Evolution of masses of the two minitori for elliptical orbit (top) and circular orbit (bottom). Red \& blue: no radiation pressure. Green \& brown: radiation pressure turned on at early stage. Orange \& purple: radiation pressure turned on at late stage. In the elliptical orbit models (top), the red, green, and orange lines represent the minitorus of the more massive SMBH}
\end{figure}

\begin{figure}
\begin{center}
\includegraphics[width=\columnwidth]{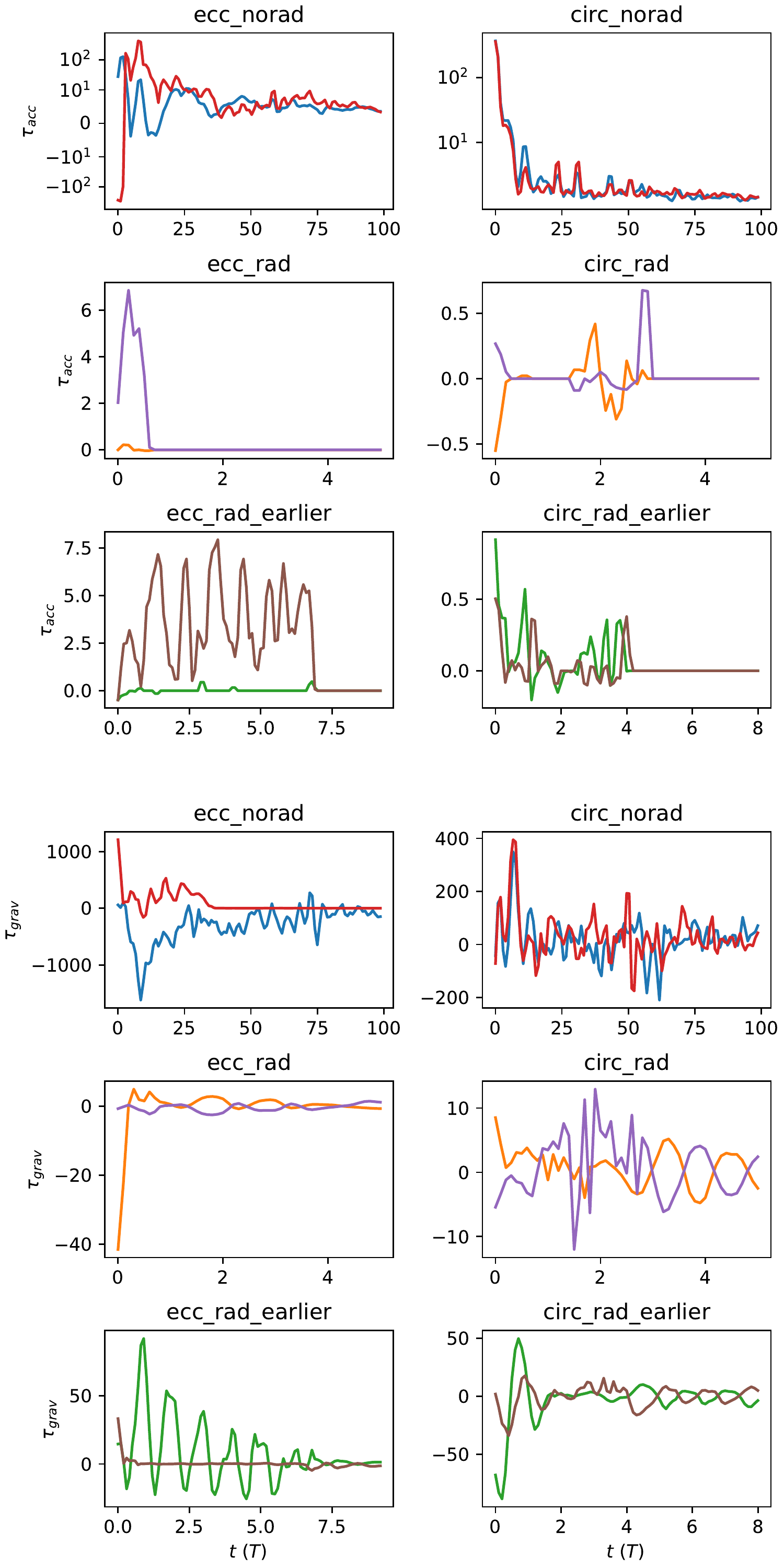}
\end{center}
\caption{\label{torque}
Gravitational and accretion torque, in all simulations, for both black holes. Units are $10^{-6} J_0/T$, as in Table~\ref{tab:meantorque}. Most plots are linear scales, but due to the large range of torque values we use a symlog scaling for ecc\_norad\_acc and circ\_norad\_acc, where $|\tau|<10^{5}$ is shown on a linear scale, and $|\tau|>10^{5}$ is shown on a log scale.}
\end{figure}

\begin{figure}
\begin{center}
\includegraphics[width=\columnwidth]{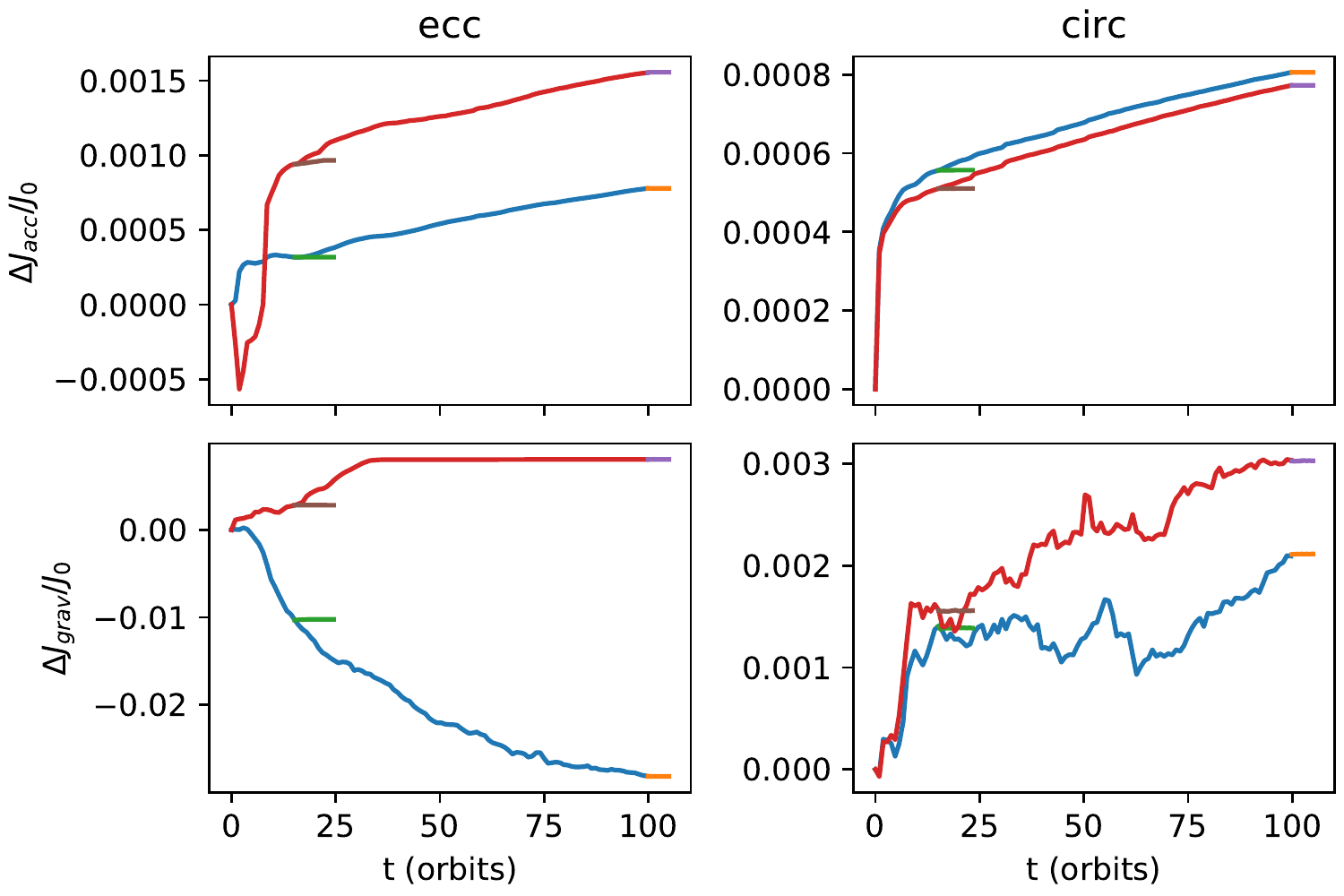}
\end{center}
\caption{\label{angmom}
Evolution of cumulative rotation impulse on the two SMBHs for elliptical orbit (left) and circular orbit (right), due to accreted angular momentum (top) and gravitational torques (bottom). Line colours are the same as in Figure~\ref{windgen}.}
\end{figure}

\subsection{Summary of morphological evolution}

Figures \ref{unary_evol}-\ref{ecc_evol} summarise the evolution of all sets of simulations.

Before radiation pressure is switched on, all models produce a flared disc, as expected as the outer disc reaches hydrodynamic equilibrium, with some spiral and ring structures in the inner regions due to hydrodynamic instabilities. As the SMBHs accrete gas particles, there are small cavities around each SMBH in every simulation, roughly on the scale of the sublimation radius.

In both binary simulations, the SMBHs carve out a larger roughly circular cavity in the circumbinary torus on the scale of the binary major axis. Spiral accretion flows extend from the inner edge of the circumbinary cavity to discs of gas around each SMBH. These discs are somewhat extended structures of cool dusty gas, much larger than the sublimation radius, and thus we refer to these structures as `minitori', to differentiate them from the ionised accretion `minidiscs'  produced by closer SMBHs binaries (see Section~\ref{sec:intro}). In the single SMBH model, the disc is continuous, and no large-scale cavity is formed.

We track the masses of the minitori in Figure~\ref{windgen}. The minitori form within about $10T$. Without radiation pressure, they lose mass through accretion on the SMBHs, and gain mass through accretion flows from the circumbinary torus. In the eccentric binary, the more massive SMBH has a much larger minitorus than the less massive SMBH, due to its larger Roche lobe. Here, gravitational stirring also causes the accretion to be quite efficient. and lower mass minitorus is completely consumed by $t=100T$, while the more massive minitorus has lost the majority of its mass. In the circular binary, both minitori only slowly lose mass, until radiation pressure is switched on.

When radiation pressure is switched on, the gas nearest the SMBHs is blown away. In the binary simulations, this leads to the destruction of the minitori. Gas from the minitori is directly driven out as winds, and radiation pressure hinders new accretion flows forming within the circumbinary cavity. All gas in the minitori is either blown away or accreted onto the SMBHs within $6T$, and the less massive minitori are destroyed within $2T$. The circumbinary flared disc is also blown outwards radially, but this has little effect on accretion processes, and does not appear to be affected by binarity as it occurs in both binary and solitary simulations.

Examples of the temperature distribution and $\rho-T$ phase diagrams of runs with and without radiation pressure are shown in Figure~\ref{tempplot}. The circumbinary torus is a cold dusty structure ($T<100$ K) and reaches the temperature floor in the plane ($T=30$ K). Without radiation pressure (and with reducing heating due to the fixed $\tau=7$), there is still some warm (up to $T\sim1000$ K) gas above and below the AGNs, due to the AGNs producing more flux in those directions. Once radiation pressure and full optical depth calculations are switched on, the circumbinary torus is pushed back. The inside of the circumbinary torus remains cold $T<100$ K), and there is a sharp transition in temperature (to $T>300$ K) at the surface where the circumbinary wind is produced. There are two effective equations-of-state of two phases present in the $\rho-T$ phase plot. At high optical depth from the AGN, the gas transitions rapidly from the temperature floor to $T=100$ K at a density of $\rho\sim10^3$ g~cm$^{-3}$. This effective equation-of-state is also present in the simulations where $\tau=7$ everywhere. At low optical depth from the AGN, there is a similar transition at $\rho\sim10^3$ g~cm$^{-3}$, but here the temperature jumps to $T=300$ K, and continues to increase as density decreases and heating becomes more efficient with respect to cooling.

\subsection{Torque}

Torque and angular impulse (i.e. cumulative torque) for all binary runs are plotted in figures \ref{torque} and \ref{angmom}. Torques are plotted from the start of each run (i.e. radiation pressure is switched on at $t=0$), while angular impulses for each orbit model are plotted together on a single time series. We only examine torques on the binary models, as we are interested in the evolution of the SMBH orbits rather than the spin of an SMBH and its accretion disc.

The systems take a little under $20T$ to equilibrate, with strong accretion \& gravitational torques during this period as the SMBHs disturb the circumbinary torus.

After settling down, and in the absence of radiation pressure, both binary models steadily \textit{gain} angular momentum through accretion from the co-rotating circumbinary torus. In the elliptical model, the less massive minitorus (red line) is destroyed by $40T$, but this is simply because the minitorus has such a low mass that the SMBH quickly accretes it. After this point, gas is accreted straight onto the SMBH particle and does not build up into a minitorus. Once radiation is switched on, at either time, accretion is drastically reduced, and angular momentum transfer from accretion effectively ceases, even when the minitorus persists for a few orbits.

Gravitational torques are stronger and more varied than accretion torques, until radiation pressure is switched on. Again, there are strong torques in the first $\sim20T$ as the systems equilibrate. In the circular model, both SMBHs then receive a net positive gravitational angular impulse, but with significant random motion. This suggests that in this case the SMBH separation would \textit{increase} and the SMBHs would not merge. In the elliptical model, the more massive SMBH receives significant \textit{negative} gravitational angular impulse, while the less massive SMBH receives some positive angular impulse until the minitorus is destroyed. In this case, provided radiation pressure remains switched off, the SMBH separation should decrease, and the SMBHs could eventually merge. However, when radiation pressure is switched on in all cases, the minitori and then the circumbinary torus are blown away, and gravitational torques become insignificant. This indicates that gravitational torques are dominated by gas within the minitori near the SMBHs, rather than gas in the circumbinary torus.

\begin{table*}
	\centering
	\caption{Mean and standard deviation of mean for both sources of torque in all binary runs. norad\_all refers to the entire runs without radiation pressure, norad\_eqm shows the mean torque after $20T$, once the system has equilibrated. Units are $10^{-6} J_0/T$ (i.e. ${\bar \tau}=1$ gives a timescale $J_0/{\bar \tau}$ of $10^6T$). Values that are $3\sigma$ significant are highlighted in bold. }
	\label{tab:meantorque}
	\begin{tabular}{rr|llll}

~&~&Ecc& ~&Circ&~\\
~&~&SMBH 1& SMBH 2&SMBH 1& SMBH 2\\
\hline
~ & norad\_all & $\mathbf{7.9 \pm 1.6}$ & $14.1 \pm 7.0$ & $9.8 \pm 4.0$ & $9.4 \pm 3.9$\\
${\bar \tau}_a$ & norad\_eqm & $\mathbf{5.49 \pm 0.22}$ & $\mathbf{6.78 \pm 0.43}$ & $\mathbf{2.833 \pm 0.095}$ & $\mathbf{3.06 \pm 0.11}$\\
~ & rad\_early & $0.0067 \pm 0.0058$ & $0.53 \pm 0.21$ & $-0.013 \pm 0.018$ & $0.027 \pm 0.019$\\
~ & rad\_late & $0.009 \pm 0.012$ & $\mathbf{2.71 \pm 0.25}$ & $\mathbf{0.062 \pm 0.018}$ & $0.026 \pm 0.012$\\
\hline
~ & norad\_all & $\mathbf{-280 \pm 30}$ & $\mathbf{86 \pm 17}$ & $20.6 \pm 7.5$ & $\mathbf{29.8 \pm 8.4}$\\
${\bar \tau}_g$ & norad\_eqm & $\mathbf{-189 \pm 19}$ & $\mathbf{44 \pm 11}$ & $10.4 \pm 6.3$ & $19.6 \pm 6.6$\\
~ & rad\_early & $-0.52 \pm 0.93$ & $\mathbf{-0.48 \pm 0.14}$ & $0.81 \pm 0.37$ & $0.58 \pm 0.66$\\
~ & rad\_late & $\mathbf{7.3 \pm 2.3}$ & $0.28 \pm 0.41$ & $-2.9 \pm 2.2$ & $-0.3 \pm 1.0$\\
\hline
	\end{tabular}
\end{table*}

\begin{figure*}
\begin{center}
\includegraphics[width=\textwidth]{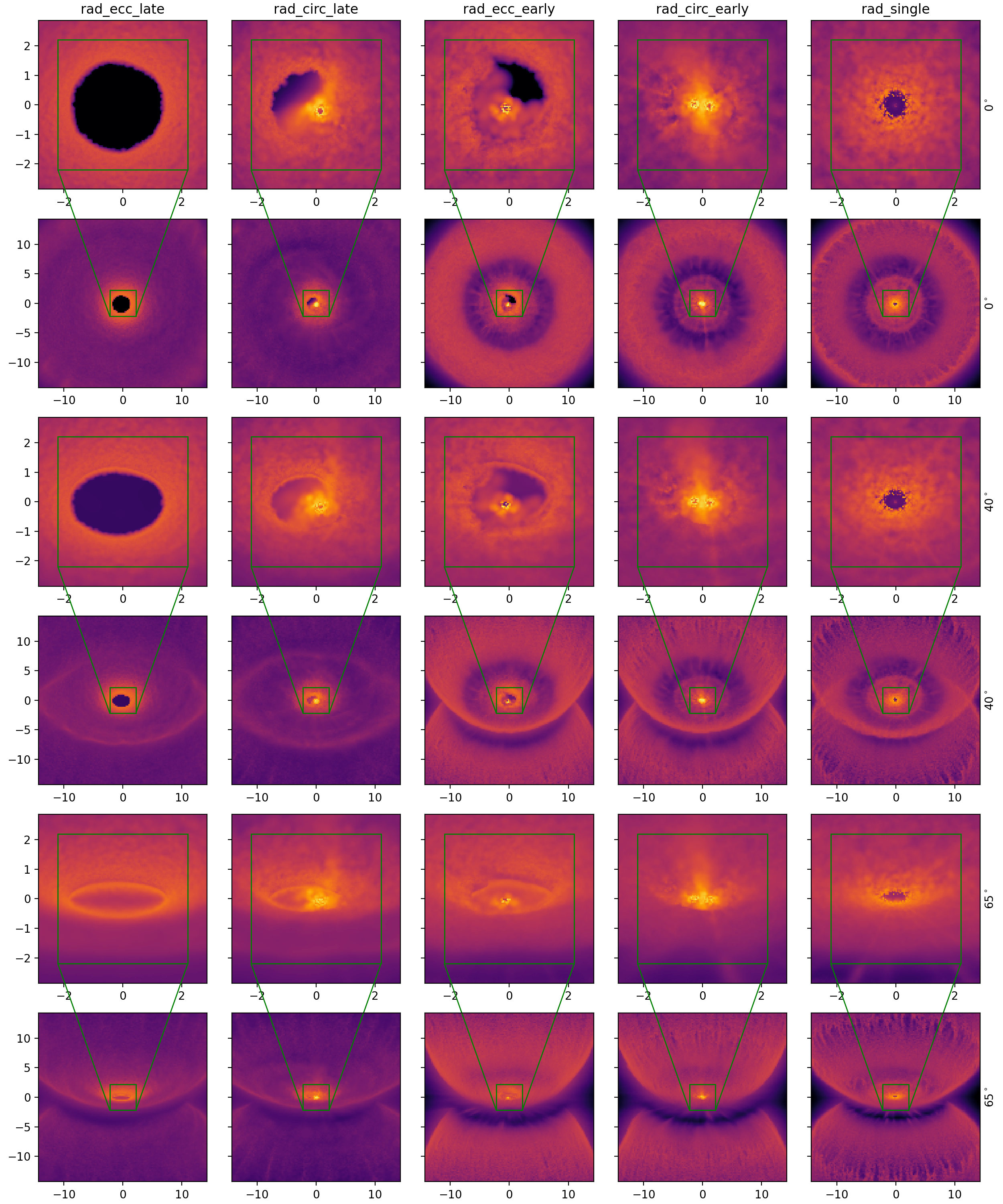}
\end{center}
\caption{\label{mockobs}
Mock observations of runs with radiation pressure at $0^\circ,40^\circ,$ and $65^\circ$ from face-on, after $2T$ of evolution.  Length units are the semi-major axis of the binary, $a$.}
\end{figure*}

The gravitational torques without radiation pressure have a large scatter (Figure~\ref{torque}). The evolution of gravitational angular impulses are potentially consistent with a random walk (Figure~\ref{angmom}). Gravitational torques may also be dominated by a small number of particles near an SMBH, where shot noise becomes a concern. Hence a more thorough statistical analysis is required to determine the applicability of our results -- i.e. whether the net sign of gravitational torques is simply the result of noise.

We calculate the mean torque $\bar{\tau}$ and the error in the mean torque in each stage of the simulation, as shown in Table~\ref{tab:meantorque}. The error of the mean is $\sigma_{\bar{\tau}}=\sigma_\tau/\sqrt{n}$, where $\sigma_\tau$ is the standard deviation of torque samples from simulation dumps. We calculate the mean torque and its error separately for accretion torque and gravitational torque, and for each black hole in each simulation. We also separate out the mean torque for $t>20T$ in the runs without radiation pressure, to show the mean torque after equilibration. By the central limit theorem, the distribution of means of a sample tends towards a normal distribution, even when the original distribution is far from normal. Hence if $|{\bar{\tau}}|>3\sigma_{\bar{\tau}}$, the mean torque is non-zero with a $\sim99.7\%$ significance. We use $3\sigma_{\bar{\tau}}$ as a threshold for significance, because with $32$ calculated means of torque, a $1\sigma_{\bar{\tau}}$ or $2\sigma_{\bar{\tau}}$ threshold would produce too many false positives.

Accretion torque is significantly positive for both orbit models after equilibration but before radiation pressure is switched on. The equilibrium phase includes strong disordered motions that wash out the torque signal. Turning on radiation pressure shuts down accretion torque in most cases, although for the smaller SMBH in rad\_ecc\_early and the larger SMBH in rad\_ecc\_late, significant positive torque persists for some time before complete blowout.

Gravitational torques are only significantly positive after equilibration for Ecc, and not for Circ. Here, the two SMBHs show significant torques of opposite sign, and the net total torque is significantly negative, and much larger than the accretion torque. This is the only case where negative torques dominate, out of all of phases of all of our runs. The binary shrinking timescale (i.e. $J_0/\tau$) is $\sim7\times10^4T$. Scaling our model to $M_t=2\times10^6\msun$, this is a time-scale of $\sim5\times10^8$~yr, which is more than sufficient to shrink the binary within a cosmological time-scale, and could potentially resolve the final parsec problem if this configuration is common.

For Circ, the gravitational torques are only $3\sigma$ significant if we include the initial equilibration phase without radiation pressure, where the out-of-equilibirum circumbinary torus produces strong torques. These gravitational torques are dominated by gas close to the SMBHs, and once radiation pressure is switched on and gas is driven out, the torques dramatically drop and are generally less signficant.

\subsection{Mock observations}

In Figure~\ref{mockobs} we plot mock infrared observations from all runs with radiation pressure after $2T$ of evolution. Infrared emissivity is assumed to be produced only by dust, and the unextinguished flux is proportional to $T_\mathrm{dust}^4$, following the Stefan-Boltzman law. Opacities are fixed at $25$ cm$^{2}$ g$^{-1}$, close to dust extinction commonly expected at K-band.

In rad\_ecc\_late,  accretion has shut off, but the binary SMBH has carved out a large hole in emission, with a diameter of $\sim2a$. The bright hot central infrared emission is therefore suppressed.

In rad\_circ\_late and rad\_ecc\_early, accretion has shut off for one SMBH, but not the other. A single bright minitorus is visible, with bright outflows. The $\sim2a$ diameter cavity is also visible. The peak emission is off-centre from the SMBH centre of mass and the centre of the cavity.

In rad\_circ\_early, both AGN are still active. Both produce outflows from their minitori, which collide in the centre, and escape perpendicularly to the SMBH-SMBH line. From edge-on views, this outflow is polar oriented, but from face-on views, it is still strongly asymmetric.

In rad\_solo, only a single outflow is generated, which only becomes polar some distance from the SMBH \citep[see][]{2020ApJ...897...26W}. The central cavity is small and circular.

\section{Discussion}

\subsection{Observational signatures of binarity, and the duty cycle of binary AGNs} \label{sec:duty}

The inclusion of radiation pressure in our dynamical modelling has provided new insight into the duty cycle of binary AGNs. The core result is that when the binary AGNs carve out a large central cavity, with a small `minitorus' around each AGN, this reduces the fuel supply to the AGNs. Gas can only accrete on to the minitori slowly through spiral accretion flows from the edge of the cavity, while gas is rapidly blown out of the minitori by radiation pressure. At this point AGN activity shuts off, and we terminate our simulation. After AGN activity is starved, the minitori should reform and AGN activity can be restarted. In a future paper we will allow the AGN luminosity to vary with accretion rate, but this is beyond the scope of the present paper. Switching on the AGN after the minitori form however does act as a simple approximation for the AGNs activating after accretion has been established.

As the minitori are low in mass compared to single torus around a single AGN, the rapid destruction of the minitori should result in an AGN duty cycle that is much more shorter than that of a single AGN. The accretion flows set up the minitori within $\sim5-10 T$, and the minitori are destroyed within $\sim1-8 T$ once radiation pressure is switched on. Scaling our simulations to a total SMBH mass of $2\times10^{6}\msun$, this is equivalent to a total on-off duty cycle period of $\sim0.5-1.5$ Myr. Scaling to $2\times10^{7}\msun$ gives a total period of $\sim0.8-2.7$ Myr.

This periodicity is longer than the observed `flickering' of quasars on $\sim10^5$ yr time-scale \citep[e.g.][]{2015MNRAS.451.2517S}, potentially caused by instabilities within the accretion disc \citep[e.g.][]{1997ApJ...482L...9S}. From observations, the long-term duty cycle of quasars ranges from $10^6-10^7$ yr \citep{2001ApJ...547...12M,2021MNRAS.tmp.1255K}. The parsec-scale binary duty cycle is in an intermediate range, consistent with the shortest quasar duty cycles, perhaps suggesting that these objects could be binaries.

The binary duty cycle is similar to the periods of modulations of accretion in simulations without radiation pressure, which range from $\sim0.5-5T$ \citep{2008ApJ...672...83M,2012ApJ...749..118S}. This is to be expected, as the time-scales of forming the minitori and of instabilities in a circumbinary disc should both be of the order of the orbital period of the inner edge of the circumbinary disc/torus.

Before blowout, our models show the classic structure of minidiscs and spiral accretion flows within a circumbinary disc, and should show the classic observational signatures \citep[e.g.][]{2019NewAR..8601525D,2019ApJ...879..110K}. Our radiation hydrodynamical simulations further demonstrate that binary AGNs should go through inactive stages as the minitori are blown out and reform. The active and inactive stages appear to have similar timescales, and so we would expect about half of binary AGNs with circumbinary gas to be active in one or both SMBHs. 

The minitori + circumbinary torus structure contains asymmetries on parsec scales, resolvable by single epoch imaging at high resolution. The best candidate instrument for this task is the GRAVITY+ interferometer on the ESO VLT \citep{gravplus}. GRAVITY recently imaged the dust sublimation region of NGC 1068 \citep{2020A&A...634A...1G}, and the increased sensitivity of GRAVITY+ will permit high resolution interferometric imaging of a large number of AGNs which can be examined for signs of these binary structures. 

\subsection{Gas Torque and the Final Parsec Problem}

The sign and strength of gravitational torques on binary SMBHs is already unclear in idealised hydrodynamic simulations of a thin disc that evolves without external perturbation or radiation pressure, given the results from smoothed-particle hydrodynamic codes \citep[e.g][]{1996ApJ...467L..77A,Escala2004,2005ApJ...630..152E,2006MNRAS.367..103D,2009MNRAS.393.1423C}, grid/mesh-based hydrodynamic codes \citep[e.g.][]{2002ApJ...567L...9A,2008ApJ...672...83M,2017MNRAS.466.1170M,2017A&A...604A.102T,2017MNRAS.469.4258T,2019ApJ...871...84M}, and semianalytic calculations \citep[e.g.][]{2008ApJ...679L..33K,2009PASJ...61...65H,2009MNRAS.398.1392L}. In those simulations, gravitational instabilities gradually grow over many orbits, producing anisotropies that may provide significant gravitational torque to the binaries. Our simulations show that the addition of radiation pressure blows away these carefully constructed anisotropies.  Additionally, on these scales we expect star formation and feedback to influence the gas \citep[e.g.][]{Wada2012}, providing a source of turbulence to smooth out any anisotropies and dampen gravitational instabilities. If the binary is small enough that the SMBHs are within each others sublimation radius, and share a circumbinary accretion disc rather than a circumbinary torus, then these effects would be less significant, but a binary would need to progress through a circumbinary torus phase to reach such a small separation.

Even without radiation pressure, we do not find a consistent torque in both sets of simulations. Strong torques are present for one orbital configuration, but not the other, indicating that these torques are very sensitive to orbit, and do not provide a universal pathway for solving the final parsec problem.

It therefore becomes increasingly unlikely that a circumbinary gas structure can robustly provide sufficient torque to cause a binary SMBH merger. However, the influence of a massive perturber such as a passing molecular cloud could perhaps provide significant torque, and we will investigate this scenario in a future paper.

\section{Conclusions}

We performed radiation hydrodynamic simulations of binary AGNs of parsec-scale separation. We investigated whether gas torques can shrink the binary, and whether we can identify new signatures for binary AGNs. We summarise our conclusions here:

\begin{enumerate}

 \item The standard morphology of minidiscs within a circumbinary disc from smaller scale simulations is still produced in our larger scale simulations, producing minitori within a circumbinary torus. Such structures could perhaps be imaged by high resolution infrared interferometers.
 \item The minitorus structures produce a shorter duty cycle in binary AGNs as the minitori are small and quickly consumed by accretion and blown away by radiation pressure.
   \begin{enumerate}
    \item The timescale is on the order of $\sim10$ binary orbital periods, of $\sim1$ Myr for a total binary mass of $2\times10^7\msun$.
    \item The active and inactive stages have similar timescales, and we should expect about half of binary SMBHs to be active
\end{enumerate}
\item We find that gravitational torques from gas can only shrink a binary within a Hubble time and solve the final parsec problem for particular orbital configurations, and only during the inactive phase of the AGN. This process is not robust or universal, and therefore parsec-scale binary SMBHs should be relatively common
   \begin{enumerate}
	\item When radiation pressure is switched off, the binary would shrink within $\sim5\times10^8$~yr and solve the final parsec problem in our elliptical orbit model, but would expand in our circular orbit model
	\item These gravitational torques mostly come from the minitori, and quickly become negligible once radiation pressure is switched on and the minitori are blown away 
	\end{enumerate}
 \item We reproduce the standard observational signatures, including asymmetries and a central cavity where hot emission is suppressed

\end{enumerate}

\section*{Acknowledgements}

This research is supported by European Research Council Starting Grant ERC-StG-677117 DUST-IN-THE-WIND and the UK Science and Technology Facilities Council (STFC) through grant ST/V001000/1. LHB is supported by DAAD (German Academic Exchange Service). 

%%%%%%%%%%%%%%%%%%%%%%%%%%%%%%%%%%%%%%%%%%%%%%%%%%
\section*{Data Availability}

Simulation data is available by request. The Python scripts used to analyse the simulations and produce the tables and plots are publicly available at \href{https://github.com/Astrokiwi/binary_torque_small}{https://github.com/Astrokiwi/binary\_torque\_small}. The repo for our modified version of GIZMO is available at \href{https://bitbucket.org/Astrokiwi/agn-rhd-model-for-gizmo-sph}{https://bitbucket.org/Astrokiwi/agn-rhd-model-for-gizmo-sph}.


\bibliographystyle{mnras}
\bibliography{binary_agn} 

%%%%%%%%%%%%%%%%%%%%%%%%%%%%%%%%%%%%%%%%%%%%%%%%%%

%%%%%%%%%%%%%%%%% APPENDICES %%%%%%%%%%%%%%%%%%%%%

%\appendix

%%%%%%%%%%%%%%%%%%%%%%%%%%%%%%%%%%%%%%%%%%%%%%%%%%

% Don't change these lines
\bsp	% typesetting comment
\label{lastpage}
\end{document}